\documentstyle[prl,aps,twocolumn]{revtex}
\begin{document}
\draft
\title{Phonon Assisted Multimagnon Optical Absorption
and Long Lived Two-Magnon States in Undoped Lamellar
Copper Oxides. supr-con/9501001 }
\author{J. Lorenzana and G. A. Sawatzky}
\address{
Laboratory of Applied and Solid State Physics, Materials Science Centre,\\
University of Groningen,Nijenborgh 4, 9747 AG Groningen, The Netherlands}
\date{\today}
\maketitle
\begin{abstract}
\widetext
We calculate the effective charge for multimagnon  infrared (IR)
absorption assisted by phonons in the parent insulating compounds
of cuprate superconductors and the spectra for two-magnon absorption
using interacting spin-wave theory. Recent measured bands  in the mid IR
[Perkins et al. Phys. Rev. Lett. {\bf 71} 1621 (1993)]
are interpreted as involving one phonon plus a two-magnon
virtual bound state,
and one phonon plus higher multimagnon absorption processes.
The virtual bound state consists of a narrow resonance
occurring when the magnon pair has total momentum close to $(\pi,0)$.
\end{abstract}
\narrowtext
The discovery of high-Tc superconductivity in doped cuprate
 materials\cite{bed86} has triggered a lot of attention in the parent
quasi-two-dimensional spin-$\frac12$  quantum antiferromagnets.
So far complementary information on the antiferromagnetism
 has come from different probes like  neutron scattering and Raman
light scattering\cite{man91}. In this work
we reinterpret recent infrared absorption measurements\cite{per93}
(Fig.~\ref{ade}) in terms of phonon assisted multimagnon absorption.
 The narrow primary peak in Fig.~\ref{ade} is explain in terms
of a long lived virtual bound state of two magnons here referred to as a
bimagnon. These new states are narrow resonances occuring when the magnon
pair has total momentum close to $(\pi,0)$ and have a rezonably well
defined energy and momentum in a substantial portion of the Brillouin zone.

In principle IR absorption of magnons  is not allowed in the tetragonal
structure of  cuprate materials. This is because in a typical two-magnon
excitation the presence of a center of inversion inhibits any asymmetric
displacement of charge and hence the associated dipole moment vanishes.
 However the  situation changes if phonons are taken into account.
In a process in which
one phonon and two magnons are absorbed, the symmetry of the lattice is
effectively lower, and the process is allowed. A similar theory was put
forward by  Mizuno and Koide\cite{miz63} to explain
magnetically related IR absorption bands found many years ago by
 Newman and Chrenko in NiO\cite{new59}. To the best of our knowledge,
we present the first explicit calculation of coupling constant
 for phonon-assisted absorption of light by
multimagnon excitations and the line shape for two-magnon absorption.
 We consider a Cu-O layer but generalization
of our results to other magnetically ordered insulators is trivial.

Consider
a  three-band-Peierls-Hubbard model\cite{yon92} in the presence of an
electric field ($\bbox{ E}$) and  in which for simplicity
Cu atoms are kept fixed and O ions are allowed to move with
displacements $\bbox{ u}_{\bbox{i+\delta} /2}$. Here $\bbox{ i}$ labels
\break
\newline
\bigskip \bigskip \bigskip \bigskip \bigskip \bigskip \bigskip \bigskip

\noindent
Cu sites and $\bbox{\delta}=\bbox{\hat{x},\hat{y}}$, so that
$\bbox{ i} +\bbox{\delta}/2$ labels O sites.

   Holes have an on-site interaction $U_d$ on Cu and $U_p$ on O, a Cu-O
 repulsion $U_{pd}$, on-site energies on Cu $E_d$, and on O $E_p$,
 and Cu-O hopping $t$ (O-O hopping is neglected here).
We define $\Delta=E_p-E_d + U_{pd}$, $\epsilon=2 (E_p-E_d)+ U_p$.
  When an O ion moves in the direction
of a Cu with displacement $|u|$, the corresponding on-site energy of Cu
changes to first order by $-\beta |u|$ and the corresponding Cu-O hopping by
$\alpha |u|$. Opposite signs apply when the O moves in the opposite
direction\cite{yon92}.

To calculate the coupling constants of light with
 one-phonon-multimagnon processes we first obtain
a low energy  Hamiltonian  as a perturbation expansion valid
when  $t<<\Delta,\epsilon,U_d$ and when the phonon field and
the electric field vary slowly with respect to typical gap frequencies,
\begin{equation}
\label{h}
H=\sum_{\bbox{i},\bbox{\delta}}
 J(\bbox{ E},\{\bbox{ u}_{\bbox{i}+\bbox{\delta} /2}\})
B_{\bbox{i}+\bbox{\delta} /2}+H_{\rm ph}-\bbox{ E} \bbox{ P}_{\rm ph} .
\end{equation}
Here $B_{\bbox{i}+\bbox{\delta} /2}=\bbox{S_iS_{i+\delta}}$ with
$\bbox{S_i}$ the spin operators, $H_{\rm ph}$ is
the phonon Hamiltonian containing spring constants and masses for
 the O ions ($M$) and $\bbox{ P}_{\rm ph}$
is the phonon dipole moment. The first term in Eq.~(\ref{h}) contains
spin-dependent fourth order corretions in $t$ whereas fourth,
second and zero order spin-independent processes
are collected in the last two terms.
 As usual\cite{esk93} we calculate the superexchange, $J$, as the
energy difference between the singlet and
triplet states of the spins located at Cu$_{\rm L}$  and Cu$_{\rm R}$ in
Fig.~\ref{con}.
 We only need to consider the three configurations (A,B,C) of the L-R bond
and $\bbox{E}$.  Next we Taylor expand
$J$ to first order in $\bbox{E}$ and
$\{\bbox{ u}_{\bbox{i}+\bbox{\delta} /2}\}$\cite{notea},
\FL
\begin{equation}
\label{j}
J=J_0+\eta (u_{\rm L} -  u_{\rm R}) - E [q_{\rm I} u_0 +
\lambda q_{\rm A} (2 u_0 -u_{\rm L} - u_{\rm R}) ]
\end{equation}
Here $\lambda=1$ for configuration A and  $\lambda=0$ for configurations
 B and C. In each configuration the displacement of the central O and the
electric field are parallel, i.e.
$\bbox{ E}=E \bbox{\hat{e}},  \bbox{ u}_0=u_0 \bbox{\hat{e}}$.
The direction of $ \bbox{\hat{e}}$ is the same as the arrows at the bottom of
Fig.~\ref{con}. $u_{\rm L}$ and $u_{\rm R}$ are only relevant in
 configuration A.
 $u_{\rm L}=   u_{\rm L1} + u_{\rm L2} - u_{\rm L3} $,
 $u_{\rm R}= - u_{\rm R1} + u_{\rm R2} + u_{\rm R3}$.
The numbering
and the direction of the displacements are shown in Fig.~\ref{con}.
The first term in Eq.~(\ref{j}) is the superexchange in the absence of
the electric and phonon fields, $J_0 =
\frac{4t^4}{\Delta^2} [\frac{1}{U_d} + \frac{2}{\epsilon}]$.
The remaining quantities are a magnon-phonon
coupling constant,
$\eta=   \frac{-4t^4}{\Delta^2}\beta [\frac{1}{\Delta}(\frac{1}{U_d} +
\frac{2}{\epsilon})+ \frac{2}{\epsilon^2}]$ and
   effective charges associated with
  one phonon and multimagnon processes,
\begin{eqnarray}
q_{\rm I}&=& -e  \frac{8t^4}{\Delta^2} [\frac{1}{\Delta}(\frac{1}{U_d} +
\frac{2}{\epsilon})+ \frac{2}{\epsilon^2}],\\
q_{\rm A}&=& -e  \frac{4t^4}{\Delta^2} \beta a_{pd} [\frac{2}{\Delta^2}
(\frac{1}{U_d} + \frac{2}{\epsilon})+\frac{1}{U_d}
(\frac{1}{\Delta}+\frac{2}{U_d})^2].
\end{eqnarray}
$a_{pd}$ is the Cu-O distance. Within a point charge estimation the
parameter $\beta a_{pd} \approx 2 U_{pd} $.
The dipole moment is obtained  as
 $\bbox{ P}=-\frac{\partial H}{\partial \bbox{ E}}$ and using Eq.~(\ref{j}) in
the  relevant configurations.
We get up to fourth order in $t$,
\begin{equation}
\bbox{ P} = \bbox{ P}_{\rm 1ph} +\bbox{ P}_{\rm 1ph+mag}
\end{equation}
The first term describes conventional phonon absorption.
We define
$\delta B_{\bbox{i}+\bbox{\delta} /2}=B_{\bbox{i}+\bbox{\delta} /2} -
\langle B_{\bbox{i}+\bbox{\delta} /2} \rangle $ and its Fourier transform,
$\delta B^{\delta}_{\bbox{ p}}$ and the Fourier transform of
$\bbox{ u}_{\bbox{i}+\bbox{\delta}/2 }$,
$\bbox{ u}^{\delta}_{\bbox{ p}}$.
After Fourier transforming, the dipole moment for one phonon and
multimagnon processes for an in-plane field in the $x$ direction
has the form,
 \begin{eqnarray}
 P_{x,{\rm 1ph+mag}}&=& N[q_{\rm I} \sum_{\bbox{ p},\delta} \delta
B^{\delta}_{\bbox{ p}}
u^{\delta}_{x\bbox{ p}}\nonumber\\
+\lambda 4  &q_{\rm A}& \sum_{\bbox{ p},\delta} \sin(\frac{p_x}{2})
\sin(\frac{p_\delta}{2}) \delta B^x_{\bbox{ p}} u_{\delta\bbox{ p}}^{\delta}]
\end{eqnarray}
and $\lambda=1$.
 The case of an electric field perpendicular to the plane
is obtained by putting  $\lambda=0$ and replacing $u^{\delta}_x$ by
$u^{\delta}_z$. $N$ is the number of unit cells.
The first term is isotropic being present in any configuration.
Looking at the cluster in Fig.~\ref{con} it can be
understood as a spin dependent correction to the charge in O$_0$.
 Its physical origin is that fourth order corrections to the charges
 involve spin dependent processes.
For example if the  spins in Cu$_{\rm L}$ and Cu$_{\rm R}$ are parallel they
cannot both transfer to  O$_0$ whereas if they are antiparallel they can.
 Fig.~\ref{pro}(a)
illustrates a typical process efficient in  configuration B.
The second term is anisotropic being present for an in-plane field only.
 It originates from a ``charged phonon'' like effect\cite{ric79}.
Consider the configuration in which the electric field and the displacement
of O$_0$ are both parallel to the Cu$_{\rm L}$-Cu$_{\rm R}$ bond
 (A in Fig.~\ref{con}),
and a phonon in which the O's around Cu$_{\rm L}$ breathe in and the
O's around Cu$_{\rm R}$ breathe out. (We don't need to consider zero momentum
phonons to couple to light since it is the total momentum,
magnons plus phonons  which
 has to add to zero.) The Madelung potential in  Cu$_{\rm R}$ decreases, and
in Cu$_{\rm L}$ it increases, creating a displacement of charge from left
to right that
contributes to the dipole moment.  Fig.~\ref{pro}(b) illustrates a typical
process. But again this effect is spin dependent since if the two spins
are parallel they cannot both transfer to Cu$_{\rm R}$.

 The real part of the
 optical conductivity due to these processes is given by
the dipole-moment-dipole-moment correlation function.
Assuming, for simplicity,  that only the $u^{\delta}_{\delta'\bbox{ p}}$'s
with the same $\delta$ and $\delta'$ mix, and decoupling the
phonon system  from the magnetic system which is valid in lowest order
in the magnon-phonon coupling we get ($\hbar=1$),
\begin{eqnarray}
\label{sdw}
&\sigma& = -\frac{\pi\omega}M \sum_{\bbox{ p}} {\rm Im}\Bigl[
\frac{q_{\rm I}^2}{\omega_{\perp\bbox{ p}}}
 \langle\!\langle \delta B^y_{\bbox{ -p}};
\delta B^y_{\bbox{ p}}\rangle\!\rangle^{\omega_{\perp\bbox{ p}}} \nonumber \\
&+&  \frac{16 \lambda^2 q_{\rm A}^2\sin^2(\frac{p_x}{2})\sin^2(\frac{p_y}{2})+
(4\lambda q_{\rm A} \sin^2(\frac{p_x}{2})-q_{\rm I})^2}
{\omega_{\parallel\bbox{ p}}} \\
& & \langle\!\langle \delta B^x_{\bbox{ -p}};
\delta B^x_{\bbox{ p}}\rangle\!\rangle^{\omega_{\parallel\bbox{ p}}}
 \Bigl] .\nonumber
\end{eqnarray}
Here $\omega_{\parallel\bbox{ p}}$ is the frequency of the
$ u_{x\bbox{ p}}^x$ and $u^y_{y\bbox{ p}}$ phonons and
 $\omega_{\perp\bbox{ p}}$ is the frequency of the
$ u^y_{x\bbox{ p}}$ and $u^x_{y\bbox{ p}}$ phonons.
$\omega_{\parallel\bbox{ p}}$ can be associated with the frequency of
Cu-O stretching mode phonons and $\omega_{\perp\bbox{ p}}$ with that of
Cu-O bending mode phonons.
The supraindex in the Green functions indicates that the poles
should be shifted by that amount.
The absorption coefficient is obtain assuming weak absorption as
$\alpha=\frac{4\pi}{c\sqrt{\epsilon_1}}\sigma$ with $\epsilon_1$ the
real part of the dielectric constant\cite{per93}(b).

To compute the magnon-magnon Green functions we
use interacting spin-wave theory\cite{ell68} with a Holstein-Primakoff
transformation. On the A (B) sublattice we put
$S^+_{\bbox{i}} (S^-_{\bbox{i}})=
\sqrt{2S(1-b_{\bbox{i}}^{\dag} b_{\bbox{i}})}b_{\bbox{i}}$ and
the corresponding Hermitian conjugates.
 $S=1/2$ in our case and $b_{\bbox{i}}$  is  a boson operator.
 With this definition the Hamiltonian is invariant
under the exchange of the sublattices and we don't need to distinguish
between them. Accordingly we work in the non-magnetic Brillouin zone.
The Heisenberg Hamiltonian in this representation is expanded to zero
order in $1/S$ and normal ordered with respect to the non-interacting
spin-wave ground state. The non-interacting part is diagonalized
by the Bogoliubov transformation
$Q_{\bbox{k}}^{\dag}=
u_{\bbox{k}}b^{\dag}_{\bbox{k}}-v_{\bbox{k}}b_{\bbox{-k}}$, where
$b_{\bbox{k}}$ is the Fourier transform of $b_{\bbox{i}}$ and
$u_{\bbox{k}}=\sqrt{(1+\omega_{\bbox{k}})/(2\omega_{\bbox{k}})}$,
$v_{\bbox{k}}=-{\rm sig}(\gamma_{\bbox{k}})
\sqrt{(1-\omega_{\bbox{k}})/(2\omega_{\bbox{k}})}$, with
$\gamma_{\bbox{k}}=2/z\sum_{\bbox{\delta}}\cos(\bbox{k}\bbox{\delta})$,
$\omega_{\bbox{k}}=\sqrt{1-\gamma_{\bbox{k}}^2}$ and
$z$ is the coordination number ($z=4$ in our case).
We define $g_{\bbox{p}\bbox{q}_1\bbox{q}_2}=
\langle\!\langle Q_{\frac12{\bbox p}+{\bbox q}_1}
 Q_{ \frac12{\bbox p}-{\bbox q}_1}
;Q_{\frac12{\bbox p}+{\bbox q}_2}^{\dag}
 Q_{\frac12{\bbox p}-{\bbox q}_2}^{\dag}
\rangle\!\rangle$.
Next we apply a standard RPA like decoupling for the
equation of motion of $g$ which can be solved
for  integrals over the Brillouin zone
 of $g$ weighted with products of $u$'s and $v$'s\cite{ell68}.
Finally $\langle\!\langle\delta B^x_{\bbox{ -p}};\delta B^x_{\bbox{
p}}\rangle\!\rangle$ can be written as a sum of such quantities with proper
weighting
factors\cite{lor94}. The dashed-doted line
in Fig.~\ref{ade} gives the contribution to the line shape
from magnon pairs with $\bbox{p}=(\pi,0)$
{\em without} the approximations that
follow.  A very sharp resonance occurs there indicating that
a virtual bound state (bimagnon) is formed.

 Analytic expressions for the Green functions can be obtained
neglecting the contributions involving $v$'s. The resulting error is small
because at short wave length $v$ is small to  start with and at long
wave lenght (low energies)  the spectral weight is small\cite{note11}.
This approximation does not shift appreciably the peak at $\bbox{p}=(\pi,0)$
and only decreases slightly the intensity\cite{lor94}.
The Green function takes the form
$\langle\!\langle\delta B^x_{\bbox{ -p}};
\delta B^x_{\bbox{ p}}\rangle\!\rangle=\frac{S^2}{N\pi}G_{xx}$ with,
\begin{equation}
\label{bb}
G_{xx}=
\frac{ G^{(0)}_{xx}+2J_0[G^{(0)}_{xx}G^{(0)}_{yy}-(G^{(0)}_{xy})^2] }
{1+2J_0(G^{(0)}_{xx}+G^{(0)}_{yy})
+4J_0^2[G^{(0)}_{xx}G^{(0)}_{yy}-(G^{(0)}_{xy})^2]}.
\end{equation}
Where,
\begin{equation}
G^{(0)}_{\delta\delta'}=\frac1{N}\sum_{\bbox q}
\frac{u_{\frac12{\bbox p}+{\bbox q}}^2 u_{\frac12{\bbox p}-{\bbox q}}^2
\cos(q_{\delta}) \cos(q_{\delta'})} {\omega- E_{\frac12{\bbox p}+{\bbox q}}
-E_{ \frac12{\bbox p}-{\bbox q}}},
\end{equation}
$E_{\bbox{k}}=E_{\rm m}\omega_{\bbox{k}}$,
$E_{\rm m}=zSJ_0(1+\zeta/2S)$ and $\zeta$ is the Oguchi
correction\cite{man91},
$\zeta\approx 0.158$ in our case.
Eq.~(\ref{bb}) takes the familiar form at $\bbox{p}=(\pi,0)$,
$G_{xx}=\frac{G^{(0)}_{xx}} {1+2J_0G^{(0)}_{xx}}$.
In this case
the imaginary part of $G^{(0)}_{xx}$
starts to be different from zero at $\omega=E_{\rm m}(\sim 2J_0)$
and rises very slowly as the frequency is increased so that it is very small
at the bimagnon energy resulting in a long life time or narrow peak.
 In the inset of Fig.~\ref{ade}
we show the imaginary part of the Green function from Eq.~(\ref{bb})
for different values of $\bbox{p}$.
This shows how the bimagnon disperses around $\bbox{p}=(\pi,0)$
it goes upwards on going towards $(0,0)$ and downwards on going towards
$(\pi,\pi)$. This indicates that $(\pi,0)$ is a saddle point and hence
it should give a Van Hove singularity when integrated over $\bbox{p}$
[Eq.~(\ref{sdw})].
The dashed curve in  Fig.~\ref{ade} is the theoretical
line shape in the same approximation
and assuming that the anisotropic processes dominate ($q_{\rm I}=0$) and
treating the phonons as Einstein like with
$\omega_{\parallel}=0.08$eV\cite{rie89}.
 It gives a surprisingly good fit for the primary peak. Such a
good fit, especially for the width was not possible within RPA
 in the Raman case\cite{sin89}. This can be partially
reconciled by the fact that a structure that is artificially
 broadened at $\bbox{ p}=(\pi,0)$ (but still much narrower than the
integrated line shape) does not change significantly the final result.

Because of the Van Hove singularity the position of the $\bbox{ p}=(\pi,0)$
bimagnon peak coincides with the peak in the line shape and is given by
$1.179E_{\rm m}+\omega_\parallel=
2.731J_0+\omega_\parallel$. This provides an alternative way to
estimate $J_0$. We found $J_0=0.121eV$ which is in good agreement
 with other estimates\cite{man91,sin89}.

 The $\omega_\parallel$ phonon is very anomalous in
orthorhombic La$_2$CuO$_4$ since it splits due to anharmonicities\cite{rie89}
its partner being at $\omega'_\parallel=$0.06eV at room temperature.
Presumably this produces the shoulder observed at lower energies in
the experiments (Fig.~\ref{ade}), although the distance to
the primary is larger than expected which may be due to
other phonons involved.  This feature was assigned to direct two-magnon
absorption\cite{per93} made weakly allowed by the lower lattice
symmetry according to the results of Ref.~\cite{tan65}.
However we found that the dipole moment for this process is directed
in the direction bisecting an angle made by the Cu-O-Cu bond\cite{mor66}
and hence can only contribute for a field perpendicular to the plane.

The oscillator strength in Fig.~\ref{ade} was adjusted to fit the experiments.
A rough estimate of the strength using $q_{\rm A}=.1e$ and
$\epsilon_1=5$ gives a value $\sim$ 4 times smaller than observed which is
quite reasonable given the uncertainties involved.

We interpret the side bands at higher energy to be due to higher
 multimagnon processes, neglected in the above approximations.
 In order to check both the validity of our approximations and the
origin of the side bands we have also computed the absorption using
exact diagonalization of a small cluster\cite{lor94}.
 The exact result confirms
the Green function calculation for the two-magnon peak and shows
also side bands corresponding to higher multimagnon processes
 which we associate with the higher energy side
bands observed in the experiments.
The relative weight of the side bands seem to be smaller
than in the experiments presumably because of
finite site effects or the presence of other processes
in the magnetic Hamiltonian as was suggested in the
 Raman case\cite{rog89}.


For an electric field polarized perpendicular to the plane only
phonons perpendicular to the Cu-O bond contribute.
We estimate the absorption to be
roughly a factor of 8 smaller than the in-plane contribution.
{}From the experimental side the anisotropy seems to be larger.
One should be aware that the cuprates are in a regime where covalency
is not small with respect to typical gap energies and hence a perturbation
in $t$ is helpful to identify the important processes and discuss
trends, but quantitative estimations are to be taken with care\cite{esk93}.
 As higher orders in $t$ are included we expect that the anisotropic
contributions grow with respect to the isotropic ones. For example
the charged phonon effects of Fig.~\ref{pro}(b) can become very efficient if
the second hole forms a Zhang-Rice singlet with the hole already present
in Cu$_{\rm L}$ since that process involves a much smaller gap.
Note also that the larger the order in $t$ the longer the range of
the processes that contribute to the anisotropic charges whereas
only local processes contribute to the isotropic charge. Longer
range processes have in general very large form factors.
These effects should give a stronger anisotropy in accordance with the
experiments. This also partially justifies our simplifying assumption
of taking $q_{\rm I}=0$.

To conclude we have computed effective coupling constants of light with
multimagnon excitations assisted by phonons and determined
 the line shape of the primary peak.  Our results explain recent
measured absorption bands in the mid IR of parent cuprate superconductors
and demonstrate the existence of very sharp virtual bound states of magnons.

We acknowledge the authors of Ref.~\cite{per93} for sending us their
work previous to publication and for enlightening discussions and
 M. Meinders for giving us the
program and helping us with the exact diagonalization calculations.
This investigation was supported by the Netherlandese Foundation for
Fundamental Research on Matter (FOM) with financial support from the
 Netherlands Organization for the Advance of Pure Research (NWO)
and Stichting Nationale Computer Faciliteiten (NCF)
Computations where perform at SARA (Amsterdam).
 J.L. is supported  by  a postdoctoral fellowship
granted by the  Commission of the European Communities.


\begin{references}
\bibitem{notea} Going to second order in
$\{\bbox{ u}_{\bbox{i}+\bbox{\delta} /2}\}$ gives processes
involving two phonons\cite{lor94} which may have been observed as
side bands in Sr$_2$CuO$_2$Cl$_2$\cite{per93}.
\bibitem{note11} The neglect of the $v$'s contributions
  produce however spurious spectral weight at low energy
 ($\omega<E_{\rm m}$). If we relax this approximation
the low energy spurious spectral weight disappears and the line shape
does not change significantly at high energies\cite{lor94}.
Accordingly we neglect this spurious contribution in Fig.~\ref{ade}.
\end{references}

\begin{figure}
\caption{Experimental data from Ref.~\protect\cite{per93}
 (solid line) and theoretical
line shape for two-magnon absorption (dashed line) in La$_2$CuO$_4$.
 The dashed doted line is the contribution to the line shape from the
bimagnon at $\bbox{p}=(\pi,0)$. The inset shows a blow-up of
 $\rho=-{\rm Im}(G_{xx})$
at the energy of the bimagnon for different values of the total momentum.}
\label{ade}
\end{figure}

\begin{figure}
\caption{Schematic representation of the cluster used in the calculations.
Full dots represent Cu's and open dots O's. Thick arrows represent the spin,
thin short arrows represent lattice displacements and thin long arrows
represent the direction of the electric field. We have represented
${\bf u}_0$ in A configuration. In general its direction is equal to the
direction of the electric field.}
\label{con}
\end{figure}
\begin{figure}
\caption{Typical processes contributing to the isotropic (a) and anisotropic
(b) effective charges. The meaning of the symbols is the same as in
Fig.~\protect\ref{con}. }
\label{pro}
\end{figure}
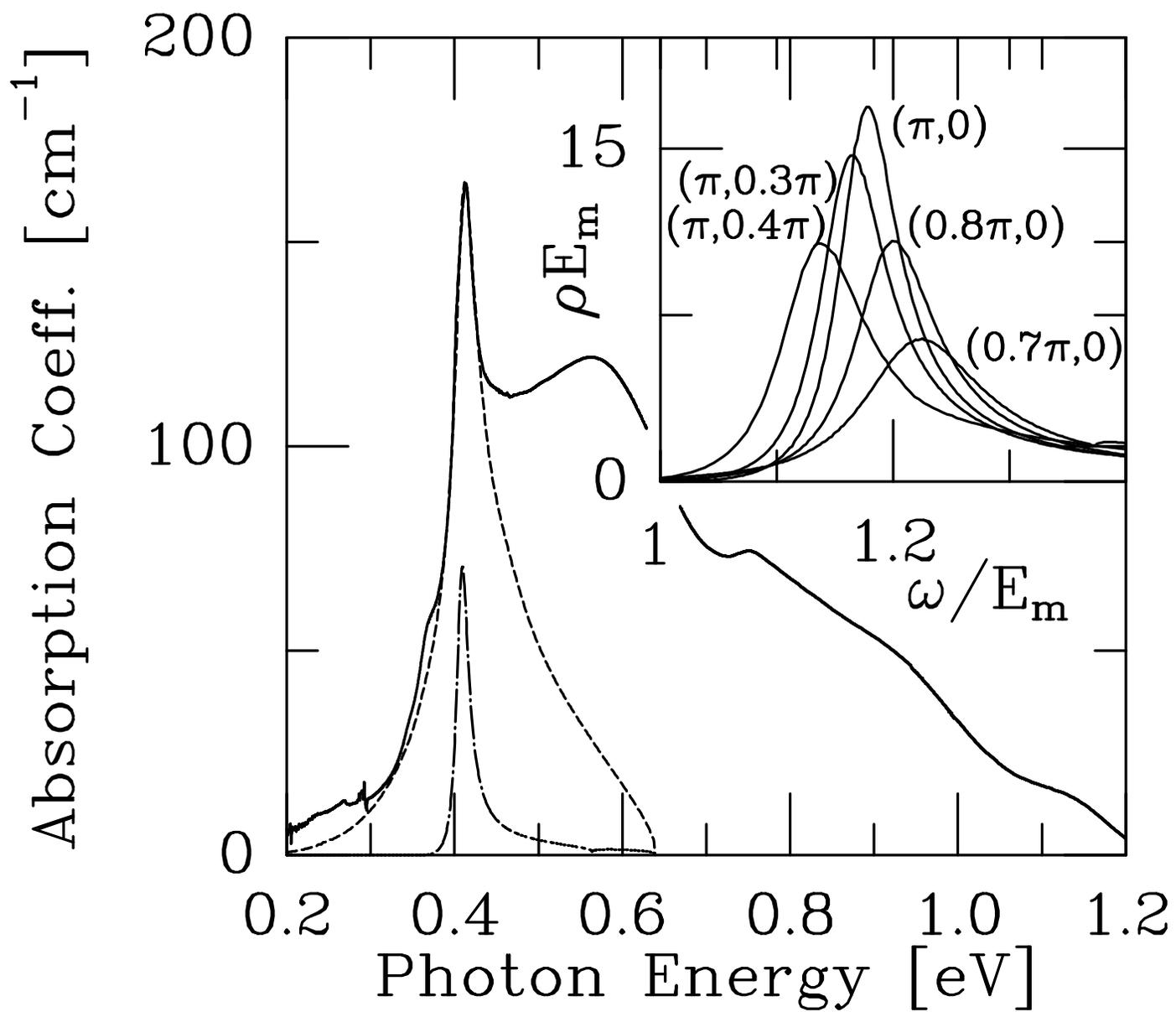
\end{document}